# Sentiment Analysis on Inflation after COVID-19

Xinyu Li[1*] & Zihan Tang[1*]

[1]School of Management of Technology, Ecole Polytechnique Federale de Lausanne, Lausanne, Vaud, Switzerland
*Corresponding Authors: xinyu.li@epfl.ch (Xinyu Li), zihan.tang@epfl.ch (Zihan Tang)

**Abstract**

We implement traditional machine learning and deep learning methods for global tweets from 2017-2022 to build a high-frequency measure of the public's sentiment index on inflation and analyze its correlation with other online data sources such as google trend and market-oriented inflation index. We use manually labeled trigrams to test the prediction performance of several machine learning models (logistic regression, random forest etc.) and choose Bert model for final demonstration. Then, we sum daily tweets' sentiment scores gained from Bert model to obtain the predicted inflation sentiment index, and we further analyze the regional and pre/post covid patterns of these inflation indexes. Lastly, we take other empirical inflation-related data as references and prove that twitter-based inflation sentiment analysis method has an outstanding capability to predict inflation. The results suggest that Twitter combined with deep learning methods can be a novel and timely method to utilize existing abundant data sources on inflation expectations and provide daily indicators of consumers' perception on inflation.

**1. Introduction**

Over the past few years, Twitter has become one of the most prominent social platforms, with 130 million users worldwide. By the end of 2020, Twitter publishes over 50 billion tweets daily on average (Note 1). Twitter service is widely used by journalists and consumers to quickly spread and obtain news in real time and has become the main information source for many users around the world. In addition, the discussions on the platform reflected the hot topics among people and revealed the collective views on political, technological, economic, and other issues. Therefore, it provides a unique opportunity for researchers studying public's beliefs. Considering the special nature of Twitter as a public forum for personal beliefs and experiences, in this study, we investigated whether Twitter conveys people's belief in short-term price dynamics and whether they can be used to trigger inflation expectations.

From the economic aspect, inflation expectations are one of the cores of any consumer and investment decisions households and businesses made. Therefore, many scholars and policymakers have carefully studied the dynamics of inflation expectations over the past years. More than that, timely and accurate understanding of inflation expectations is crucial to monetary policy, because longer-term inflation expectations can measure the credibility of central banks, while shorter-term inflation expectations reflect the effectiveness of monetary policy (Note 2). There is a commonly used source of inflation expectations: prices of financial assets linked to inflation. In this paper, we analyze the spread of price of treasury bonds traded in the market excluding inflation and including inflation. These statistics are readily available at high frequencies but are imperfect measures of consumers' inflation expectations. Indeed, they reflect investors' inflation expectations and time-varying risk premium.

To address this issue, many researchers have explored whether the vast number of metadata and documents freely available in online resources can be used to track and forecast economic variables. With the advent of the Internet and digital platforms, people have fast-growing habit of searching on the Internet and expresses opinions and emotions in social media and digital platforms. Some researchers have checked whether Internet resources available in a more timely and frequent way than traditional data can be used to assess the expectations of economic variables. There are various online resources such as social media Twitter, internet search engine Google. They provide various statistics and digital or printed documents uploaded by government agencies, academic institutions, decision makers and regulators, etc. Each of these alternative resources has been experimentally evaluated by researchers for different purposes. For example, Guzman (2011) explored the usefulness of Google search data on tracking/now casting various economic activities and macroeconomic variables; Agarwal et al. (2011) attempted sentiment analyses based on Twitter messages.

In this paper, we propose the Google and Twitter as the primary source of information for capturing consumer inflation expectations. We find the Google trend indexes of inflation-related keywords and collected weekly data for a period of five years from January 2017 to December 2021. Besides this, similar keywords-selecting methods are applied to Twitter tweets for the same period and daily data on these resources are collected. These data can be timely because Twitter messages and Google trend indexes are constantly updated. Given the wide and diverse user bases of Google and Twitter, it provides accurate information about expected inflation rates for a large sample of consumers and is not affected by risk premium constraints. If successful, this approach can complement existing data sources on inflation expectations and provide daily and weekly indicators of consumer beliefs on inflation.

The rest of the paper will be organized as following: In Section 2 — Literature Review, we briefly summarized relative literature inspiring our own thoughts. Then, comparison and explanation on the methods we used to handle our data and select models will be provided in Section 3 - Methodology, and the process how we calculate sentiment index with selected model will be discussed in Section 4 — Inflation Sentiment Index. In Section 5 — Sentiment and Inflation, the result model and relationship we find between the sentimental index and inflation will be showed and analyzed.

## 2. Literature Review

When doing sentiment analysis on people's expectation of inflation, we would like to make sentiment classification, which has been performed for blogs, comments, and micro-blogs. Due to the word limit, such texts do not contain full sentences and even have abbreviations and noisy contents. Therefore, more intelligent sentiment analysis methods are needed.

*2.1 Lexicon*

Basic lexicon methods rely on two main approaches: "bag of words" and "semantic orientation". The first attempts to build a positive/negative document classifier based on occurrence frequencies of the various words in the document, while the other classifies words (usually automatically) into two classes, "good" and "bad", and then computes an overall good/bad score for the text. Nasukawa and Yi (2003) created a sentiment lexicon of 3513 sentiment terms with consideration of the syntactic dependencies among the phrases and subject term modifiers. One of the most famous dictionary-based methods is VADER which scored the sentiment based on the dictionary and grammar rules (Hutto and Gilbert, 2014). However, Whitelaw et al. (2005) pointed out that the "atomic units" of expressions are not individual words, but appraisal groups. Lu et al. (2010) also noticed that strength of sentiment could be affected by strength of adverbs. Therefore, making adjustment on opinion words based on score of adverbs could have better result. What's more Deng et al. (2014) considered the importance of a term and its importance to express the sentiment in a document, which changed traditional unsupervised lexicon method to supervised learning method. As for features constructed from lexicon methods, Li and Xu (2014) used the emotions cause extraction technique to help in removal of unnecessary features. They also employed chi-square method to remove irrelevant features.

*2.2 Machine Learning*

Pand et al. (2002) pioneered in applying machine learning methods in sentiment analysis, such as Naive Bayes (NB), Maximum Entropy (ME) and Support Vector Machine (SVM). For NB algorithm, Rennie et al. (2003) proposed complement naive Bayes (CNB) algorithm, which is an adaptation of the standard multinomial naive Bayes (MNB) algorithm and can provide more stable results. Kang et al. (2012) pointed out that the positive classification accuracy and the negative classification accuracy did not achieve similar levels and proposed an improved Naive Bayes to achieve higher accuracy. For SVM, Li and Li (2013) argued that opinion subjectivity and expresser credibility should also be taken into consideration. They made experiments on twitter posts and found out that user credibility and opinion subjectivity is essential for aggregating micro-blog opinions. As for ensemble methods, Wang et al. (2014) compared three ensemble methods: bagging, boosting and random subspace based on NB, ME, Decision Tree (DT), K-Nearest Neighbor (KNN) and SVM for sentiment analysis. They reported better performance of ensemble methods over base learners. Moraes et al. (2013) made comparison between NB, ME and Artificial Neural Network (ANN) and found that when features were too large, feature selection method did not help gain accuracy. Also, in many cases, researchers could use more than one kind of classifiers. Bai (2011) presented a two-stage prediction algorithm. The first-stage classifier learned conditional dependencies among the words and encoded them into a Markov Blanket Directed Acyclic Graph for the sentiment variable. In the second stage, she used a meta-heuristic strategy to fine-tune their algorithm to yield a higher cross-validated accuracy.

*2.3 Deep Learning*

Up until early 2000s, the study of deep learning has never gained much popularity due to high computational costs. However, with the emergence of more powerful computers and abundant amount of data, DL methods have become the top-of-the-art technique in various domains such as sentiment analysis. For example, dos Santos and Gatti (2014) proposed a new deep convolution neutral network (CNN) which contained two layers and was designed to capture features on a from-character-to-sentence level. CNNs were a class of artificial neural networks that gained prominence with vision recognition tasks and significantly outperform traditional ANNs. In order to make up for CNNs inability to effectively interpret time information, recurrent neural networks were developed on the base of work by Rumelhart et al. (1986). However, RNN structure was unable to train gradients for longer series of inputs which led to the vanishing and exploding gradient problem. Hochreiter and Schmidhuber (1997) noted the drawback of RNNs and presented long short-term memory (LSTMs) in which the repeating cell has three gating mechanism compared with only one in RNNs. LSTMs have proved to be effective for longer series of inputs. Cho et al. (2014) proposed gated recurrent unit (GRU) which inherited the pros of LSTMs but possessed a simpler internal architecture. An influential paper by Bahdanau et al. (2015) raised attention mechanisms to spotlight in NLP. Attention mechanisms allowed the model to exploit the most relevant parts of the input sequence with flexibility. Devlin et al. (2019) put forward a transformer-based machine learning technique for NLP called BERT. Upon publication, BERT achieved outstanding performance on natural understanding tasks such as GLUE, SQuAD and SWAG. Rogers et al. (2020) concluded that nowadays it has become a universal baseline in NLP experiments. This method was also selected in this paper for implementation after evaluation.

## 3. Methods

*3.1 Data Acquisition*

3.1.1 Twitter

To analyze the people's sentiment, we use the message data from social media platform Twitter. To build our database, following methods of Angelico et al. (2022), we created an initial dictionary containing keywords related to topic of price, inflation and deflation. The dictionary is constructed in the following way:

- Price: "price", "prices", "food price", "gasoline price", "gas price", "rent price". Those words can capture tweets about prices of necessities while without much information on price dynamics.
- Inflation: "inflation", "hyperinflation", "high gasoline price", "high food price", "high gas price", "high rent price". Those words directly relate to topic of inflation or the increment of prices of necessities.

- Deflation: "deflation", "disinflation", "low gasoline price", "low food price", "low gas price", "low rent price". Such words are direct description of deflation of the decreasing price level of necessities.

Then we acquired the tweets data through academic API of twitter. Table 1 shows the basic information of our data.

Table 1. Key indicators of Twitter data

| Indicators | Values | |
|---|---|---|
| Time Period | 2017 - 2021 | |
| Number of Tweets | over 140000 | |
| Number of Users | over 68000 | |
| | US | 44.72% |
| | GB | 15.33% |
| Locations | EU | 5.01% |
| | CA | 4.76% |
| | ASIA | 5.08% |

From Table 1 we can observe that there are huge differences on active level of users across countries. Most tweets with location tags come from U.S. and U.K. which are also the center of economy.

Figure 1 shows the distribution of the number of tweets in different areas from 2017 to 2021. We can find that the United States and the United Kingdom take most part of the tweets, followed by Canada, European Union and Asia main economic bodies (China, India, Japan, Korea, Malaysia, Thailand and Singapore). Therefore, it is reasonable for us to take U.S., U.K. and total global tweets as examples to analyze in the following sections.

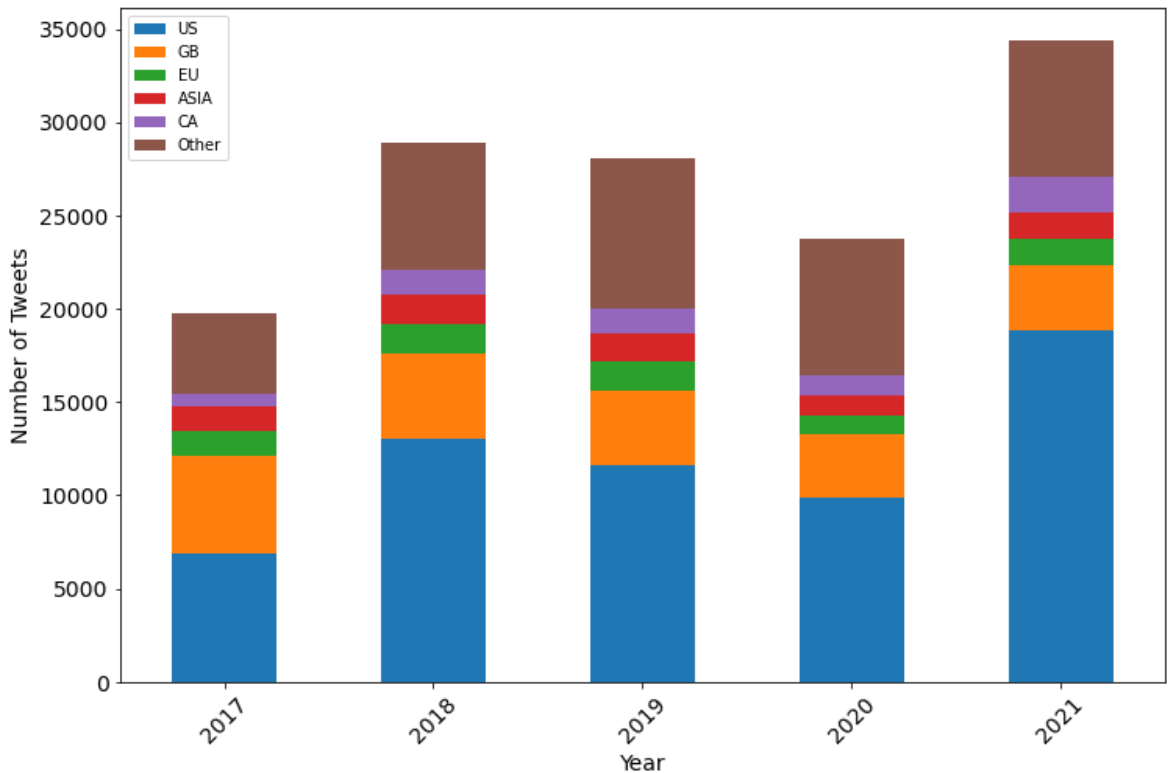

Figure 1. Time trend of number of tweets of different countires and countries

For model training and selection, we use data set of Mendelay Data (Note 3), which is composed of tweets and three sentiments labels, negative, neutral, and positive. There are 72250 tweets labeled positive, 55213 neutral and 35510 negative tweets in total. We randomly split the data set into 80% training set and 20% validation set. Also, to select model based on generalization ability, we randomly selected 10% of the tweets described in Table 1 and manually labeled the sentiment score with 1 as positive, 0 as neutral and -1 as negative.

3.1.2 Google Search

Besides the twitter data, we also want to assess the inflation via the Google keyword search index. The dataset consists of two main components: One are the official statistics on the measures of inflation rate including consumer price inflation, producer price inflation and the House Price Index, which form the basis of estimating inflation. All these series of data are collected monthly for a period of five years from Jan 2017 to Dec 2021. The other are the google trend index for selected keywords related to inflation, prices, and price dynamics. We choose the data on global, U.S. and U.K respectively. The dictionary of selected keywords in English can be categorized as follows:

- Neutral words: "price", "cost of living", and "interest rate" capture the trend about prices in general that do not provide information on price dynamics unless further analyzed.
- Positive words: "inflation", " expensive bills", "high materials prices", "high gasoline prices", "high rent" and "high house prices" reflect certain price dynamics and capture trend about increasing prices.
- "Deflation", "disinflation", "promotions", "sales", "low cost of livings" and "less expensive bills" reflect the trend about decreasing prices.

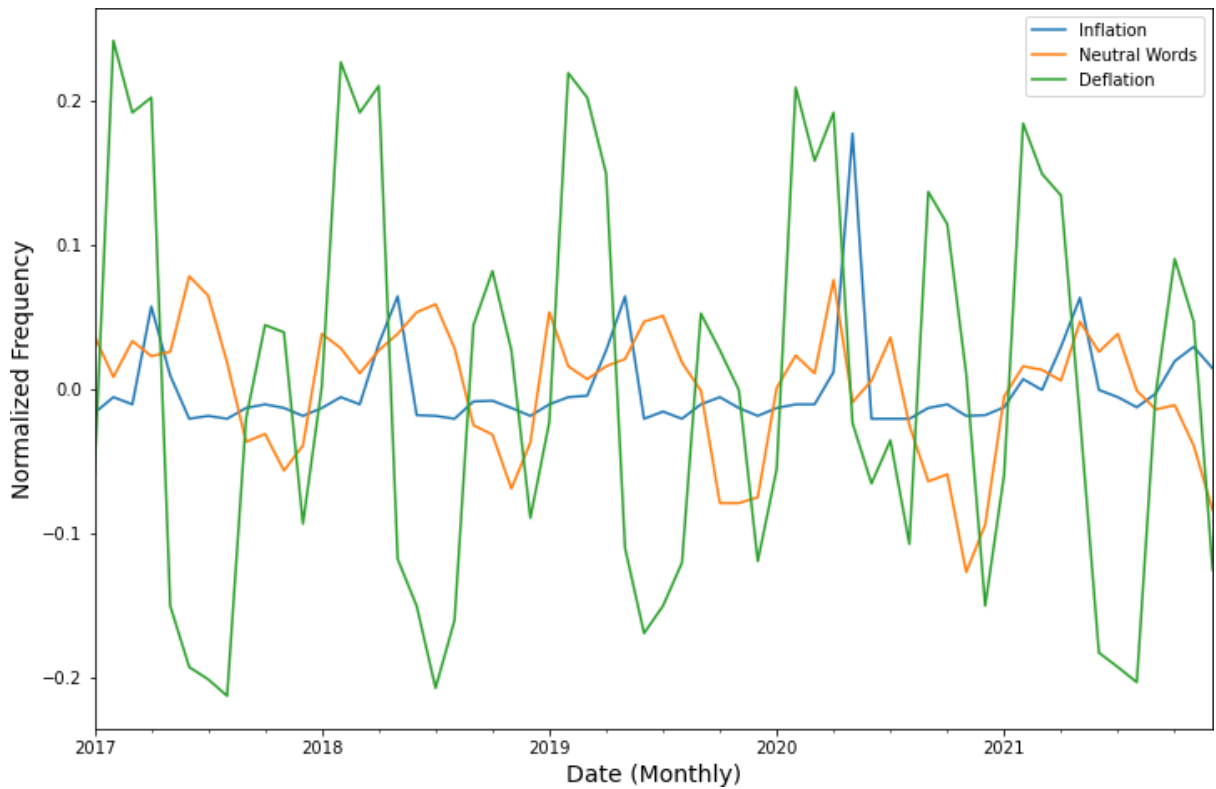

Figure 2. 2017-2022 U.S. Google Trend Search frequency of Inflation-related Words

From Figure 2, we can see that the search popularity of these terms in US are all cyclical over this 5-year period and the cycle is about 50 weeks (one year). For the inflation words, the Google Trend search quantities fluctuatingly increases over the first half of year and peak at the middle of year. Then, it consistently declines to the lowest at the end of year. Additionally, we find that Google Trend searches for deflation words increase significantly at the beginning of each year. Then experience a significant decline around the middle of the year. After a small increase, it reaches the lowest search index at the end of year. However, the search quantities of neutral words are relatively steady except of a suddenly peak in the middle of every year. That means American people care more about inflation-related things at the first half of year, especially at the middle of the year and cares less at the end of year.

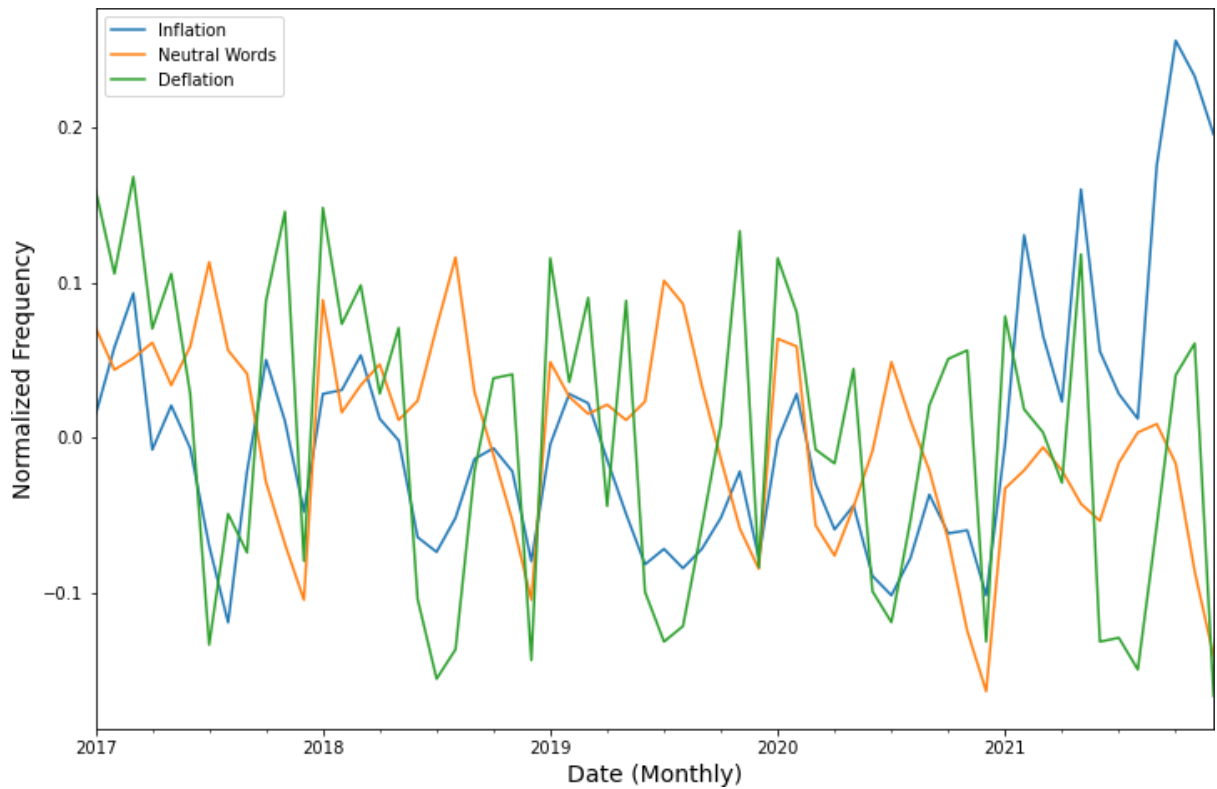

Figure 3. 2017-2022 U.K. Google Trend Search frequency of Inflation-related Words

From Figure 3, we can see that the search popularity of these terms in U.K. are all asymptotically cyclical over this 5-year period and the cycle is about 50 weeks (one year). For the inflation and deflation words, their Google Trend search fluctuated similarly. They peak at the beginning of the year and suffers a significantly decline over the first half of year and reaches the lowest at the middle of year. Then, it increases over the third Quatre and finally plummet to the second lowest point at the end of year. However, the search quantities of neutral words are relatively stable over the first half year and peak at the middle of every year. Finally, it consistently decreases to the lowest point at the end of the year. This suggests that Britons' interest in inflation-related matters fluctuates wildly over the course of the year. They were very concerned about inflation and deflation related topics in the first and third quarters, but not in the middle and end of the year.

Figure 4 indicates that the search popularity of these terms in global scale all also cyclical over this 5-year period and the cycle is about 50 weeks (one year). For the inflation words, the Google Trend search quantities fluctuatingly increases over the first half of year and peak at the middle of year. Then, it consistently declines to the lowest at the end of year. Additionally, we find that Google Trend searches for deflation words increase significantly at the beginning of each year. Then experience a significant decline to the lowest point at the middle of the year. After an obvious increase, it declines again to the second lowest search index at the end of year. However, the search quantities of neutral words are relatively steady except of a suddenly peak in the middle of every year. That means people in the world care more about inflation-related things at the first half of year, especially at the middle of the year and cares less at the end of year. And they care deflation-related things more at the first and third quatres, but less at the mid and end of the year.

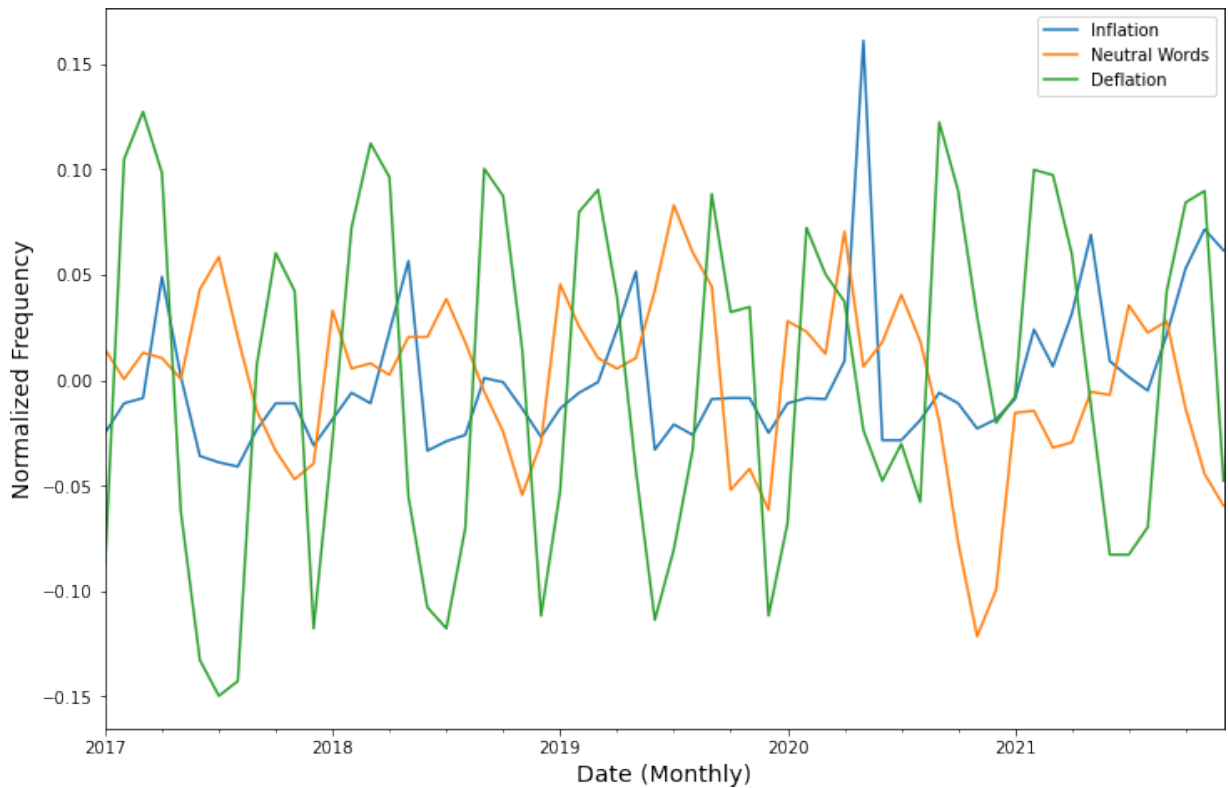

Figure 4. 2017-2022 Global Google Trend Search frequency of Inflation-relatedWords

Table 2 gives the frequencies of Google trend in U.S., U.K., and global.

Table 2. Frequencies of Google Trend in Different Regions (2017 - 2021)

| Country | | Frequency (%) | | | | |
| --- | --- | --- | --- | --- | --- | --- |
| | | Before COVID-19 | | | After COVID-19 | |
| | | 2017 | 2018 | 2019 | 2020 | 2021 |
| US | Inflation | 18.06 | 18.68 | 18.56 | 21.38 | 23.32 |
| | Neutral | 20.56 | 20.16 | 19.91 | 19.45 | 19.93 |
| | Deflation | 19.78 | 20.07 | 19.84 | 20.90 | 19.41 |
| UK | Inflation | 20.22 | 19.33 | 18.50 | 18.07 | 23.88 |
| | Neutral | 20.98 | 20.45 | 20.32 | 19.14 | 19.11 |
| | Deflation | 21.60 | 19.64 | 19.71 | 19.83 | 19.22 |
| Global | Inflation | 17.94 | 18.81 | 18.81 | 20.30 | 24.13 |
| | Neutral | 20.36 | 19.99 | 20.24 | 19.72 | 19.69 |
| | Deflation | 19.76 | 19.92 | 19.80 | 20.25 | 20.27 |

3.1.3 Market-oriented Statistics

One general method of measuring inflation expectations lies within the treasury market. From this perspective, we consider two types of bonds as measure of inflation expectation on the market: one is 5-year constant maturity treasury securities, the other one is called "inflation-indexed" 5-year bond which incorporates realized inflation into the yield. Excluding the real yield from the nominal one, we obtain a break-even inflation rate which reflects market's expectation of future inflation.

Table 3 demonstrates the descriptive statistics of break-even inflation rate in U.K. and U.S. and Figure 5 gives a more visualized version. We observe similar trend of break-even inflation rate in both countries. We notice that there is a sharp drop in inflation in both countries from February to April in 2020 and this is mainly due to the drop of energy inflation caused by falling oil price Matuszewska-Janica et al. (2021). The difference of magnitude of inflation impact is due to the contribution of energy inflation in the aggregate inflation. From then, there is an evident upward trend for inflation. There are three temporary factors contributing to the phenomenon. The first factor is "base effect" which is recovery from the changes in inflation during the last few months. The second factor is supply chain disruption. Increase in cost of production is contagious and spreads to all sectors of supply chain and eventually results in inflation on the consumer side. The last factor is pent-up demand. Previous customer anxiety and public health restrictions suppress demand for services such as hotels and restaurants. With the undoing of quarantine and more people getting vaccinated, demand begins to surge and fuels inflation. Economists name it demand-pull inflation.

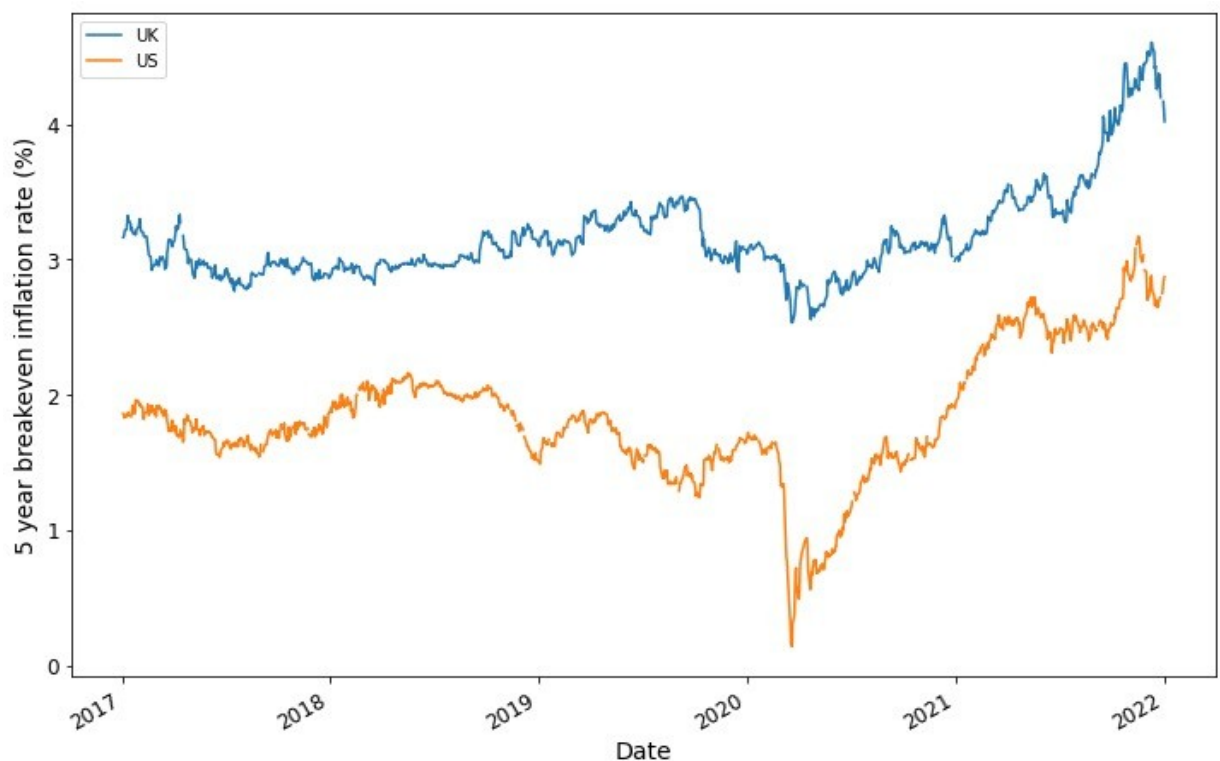

Figure 5. 5-year Break-even Inflation Rate of U.S. and U.K.

Market-oriented method tends to have higher updating frequency which proves to be more informative. Given its market orientation feature, it also produces a timelier and more realistic reflection of market participants' expectation. However, we should also reserve discretion for the creditability of the market-oriented model. This model contains other inseparable risk premiums, especially when the market exhibits high volatility and uncertainty. The model could deviate from the real market inflation expectation under this scenario. Given this limitation, stripping non-inflation risk premium and focusing directly on people's expectation of future inflation on a relatively high frequency is a possibly feasible solution. This is also what we expect to obtain from the social media approach.

Table 3. Descriptive Statistics on Inflation in UK and US

| Country | Stats (%) | 2017 | 2018 | 2019 | 2020 | 2021 |
|---|---|---|---|---|---|---|
| US | mean | 1.74 | 1.97 | 1.60 | 1.33 | 2.55 |
|    | std  | 0.10 | 0.13 | 0.16 | 0.41 | 0.22 |
| UK | mean | 2.98 | 3.01 | 3.22 | 2.96 | 3.66 |
|    | std  | 0.13 | 0.10 | 0.14 | 0.18 | 0.43 |

*3.2 Preprocessing*

For twitter data, there are frequently four components in the tweets: users, links, topics and texts. Therefore, regarding each part, we applied different approaches to process the raw tweet data.

- users: the usernames in tweets are started with '@', so we just change the username to '@user' to avoid the potential effects brought from the positive/negative words in usernames.
- links: some tweets contain links started with 'https', and the words in the links may contain words with sentiments, so to eliminate noises, we will convert all the link strings to 'https'.
- topics: the topics in the tweets start with '#'. However, topic words are often components of the whole sentences. Therefore, when dealing with the topic, we just drop the '#' and keep the topic words.
- emojis: due to the time limit and consideration of different using patterns in the emojis across different countries and areas Kejriwal et al. (2021), we simply drop the emojis from the text to reduce the potential mislabeling of the emoji.
- slangs: for slangs in the tweets, we do not make further processing for them, because according to Derczynski et al. (2013), here are noises in the slangs and inappropriate methods will lead to more noise. Therefore, to avoid further losses, we do not make processing.
- texts:
  – punctuation: we simply eliminate punctuation from the tweets, even though we may suffer from loss of information carried by certain punctuation such as '!'.
  – stop words: eliminate the stop words such as 'a', 'and' and 'the' which can appear in high frequency but have little impact on the sentiment.
  – stemming: for machine learning method, we apply Snowball stemmer (Note 4) to group the words with the same root. What's more, Snowball algorithm can perform better on the short string.

*3.3 Model Selection*

For model selection, we applied different methods extracting features from the tweets (VADER, TF-IDF and BERT). We also tried different learning algorithms (logistic regression (LR), naive Bayes (NB), complement naive Bayes (CNB), support vector machine (SVM), random forest (RF), gradient boosting tree (GBT) and BERT). Noted that for the BERT model, we choose BERT-BASE model (Devlin et al., 2019). The performance of each method on the training set and validation set has been shown in Table 4:

Table 4. Comparison of Different Models

| Feature | Model | Train (%) | Valid (%) | FP (%) | FN (%) | Test (%) |
|---|---|---|---|---|---|---|
| VADER | LR | 59.75 | 59.76 | 30.75 | 8.57 | 57.62 |
|  | NB | 57.40 | 57.25 | 22.17 | 8.11 | 41.90 |
|  | SVM | 59.50 | 59.88 | 44.36 | 7.63 | 54.43 |
|  | RF | 79.72 | 58.80 | 27.64 | 18.90 | 58.88 |
|  | GBT | 61.68 | 61.51 | 30.34 | 10.55 | 61.68 |
| TF-IDF | LR | 79.29 | 43.94 | 33.20 | 21.05 | 37.93 |
|  | CNB | 72.05 | 37.93 | 34.51 | 32.20 | 27.99 |
|  | RF | 98.02 | 41.91 | 31.82 | 21.82 | 30.99 |
| BERT | BERT | 98.40 | 93.30 | 0.00 | 0.00 | 79.99 |

Table 4 describes the performance of different models, that column Train and Valid describe the performance of the model on training set and validation set, while column Test corresponds to the model performance on out-of-sample data. From the results we can reach the following conclusion:

- Machine learning methods have bias on unbalanced data set. For all machine learning methods, we observe that they have much higher false positive rate (FP) than false negative rate (FN) on validation set, indicating that because of larger amount of positive labeled tweets.
- Encoding methods have more impact on the result then the model. From Table 4 we observe that under the same tokenizing method, there are little differences among the performance different models, while different tokenizing methods will lead to significant performance among models.
- TF-IDF tokenizing methods can lead to over-fitting problems for machine learning methods. Because tweets are always short and the words of tweets are widely spread, the matrix of TF-IDF is very large and sparse so simple machine learning cannot "learn" well.
- BERT tokenizer and model can successfully learn the patterns of the tweets with little out-sample error.

For the comparison among VADER, TF-IDF and BERT, we make the following conclusion:

- TF-IDF is a bag-of-words model. Because tweets are relatively short while contain as many kinds of words as long articles, TF-IDF methods will create a sparse feature matrix, which will result in overfitting problem.
- VADER: VADER is a lexicon-based method. We used manually labeled sentiment words to calculate the score of negative scores, positive scores and neutral scores of tweets. Therefore, for each tweet, we only get three features which is too small for the texts.
- BERT: according to Devlin et al. (2019), BERT model also takes the relative location of each word in the text, which is not considered in TF-IDF and VADER. What's more, BERT transformer is pretrained on the huge data base, indicating that it can better deal with the small-scale data.

Therefore, we will use BERT method to analyze the tweets data from 2017 to 2021. Also, the pretrained parameters obtained from the training set will also be used to calculate sentiment score of the tweets data we acquired through API.

## 4. Inflation Sentiment Index

*4.1 Index Construction*

To construct sentiment index, we calculate average sentiment score in each day, which indicates people's overall attitudes towards the inflation. What's more, we also notice that frequency of the tweets in each day also varies a lot in different episodes. Therefore, we plot the summary description of the sentiment index (global, U.S. and U.K.) and the global frequency in Figure 6:

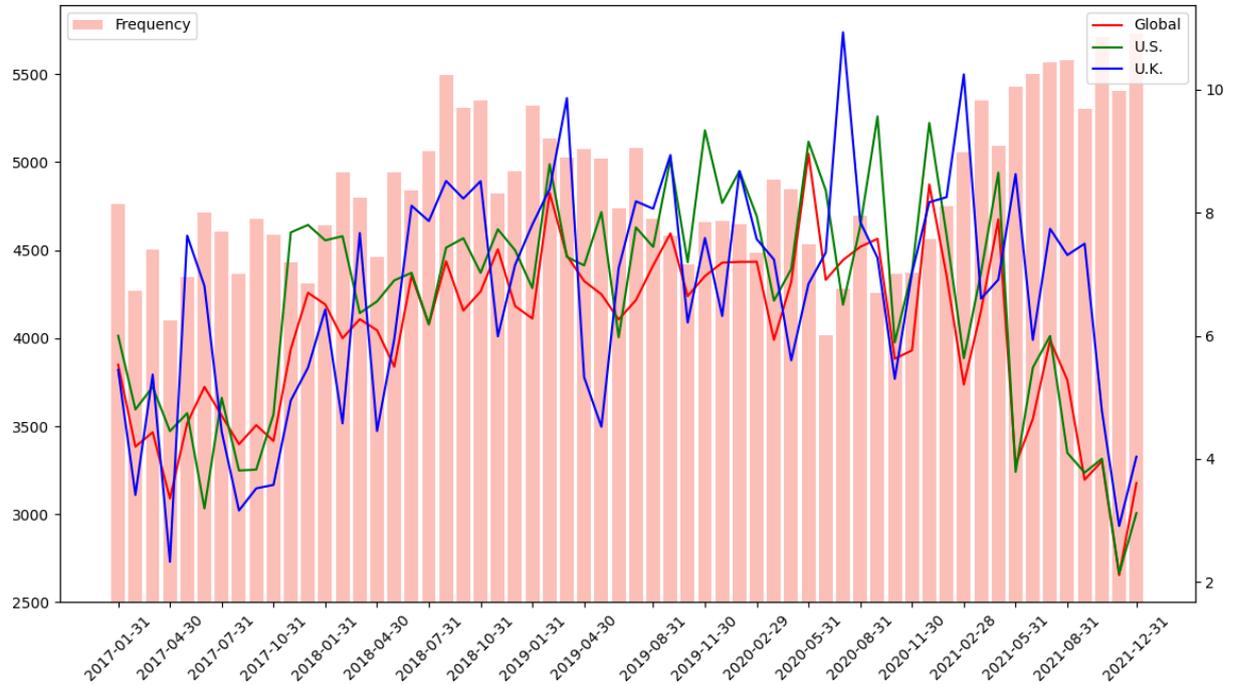

Figure 6. Global Sentiment Score

We also observe that rolling volatility of sentiment score also varies a lot through time and among countries. We calculate the 30-day rolling volatility of sum of total sentiment score each day in Figure 7. The summary statistics of the index is also shown in Table 5.

Table 5. Summary Statistics of Sentiment Index

|  | mean | std | max | min |
| --- | --- | --- | --- | --- |
| Freq | 80.32 | 40.64 | 499.00 | 4.00 |
| Score |  |  |  |  |
| Global | 0.21 | 0.11 | 0.80 | -0.28 |
| U.S. | 0.23 | 0.17 | 0.81 | -0.57 |
| U.K. | 0.22 | 0.31 | 1.00 | -1.00 |
| M.Vol(30) |  |  |  |  |
| Global | 10.17 | 3.51 | 27.81 | 4.88 |
| U.S. | 6.12 | 2.94 | 21.67 | 2.74 |
| U.K. | 3.55 | 1.15 | 7.99 | 1.71 |

*4.2 Patterns in different periods*

We would like to divide the time into three episodes: before COVID-19 (2017 – 2020), during COVID-19 (during 2020) and after COVID-19 (2021) to analyze patterns of people's sentiment in different episodes and how people's sentiment towards inflation after COVID-19.

To further analyze people's sentiment towards the inflation in different periods, we apply auto regression (AR) model with time trends to represent people's expected sentiment indifferent episodes. The coefficient of time trend $t$ represents people's long-term attitudes in one period. We applied ACF (autocorrelation function) and PACF (partial autocorrelation function) to choose parameters. Figure 8, Figure 9 and Figure 10 show the fitted results (the orange, red and brown lines represent the fitted value in each period respectively), and the result of regressions are reported in Table 6.

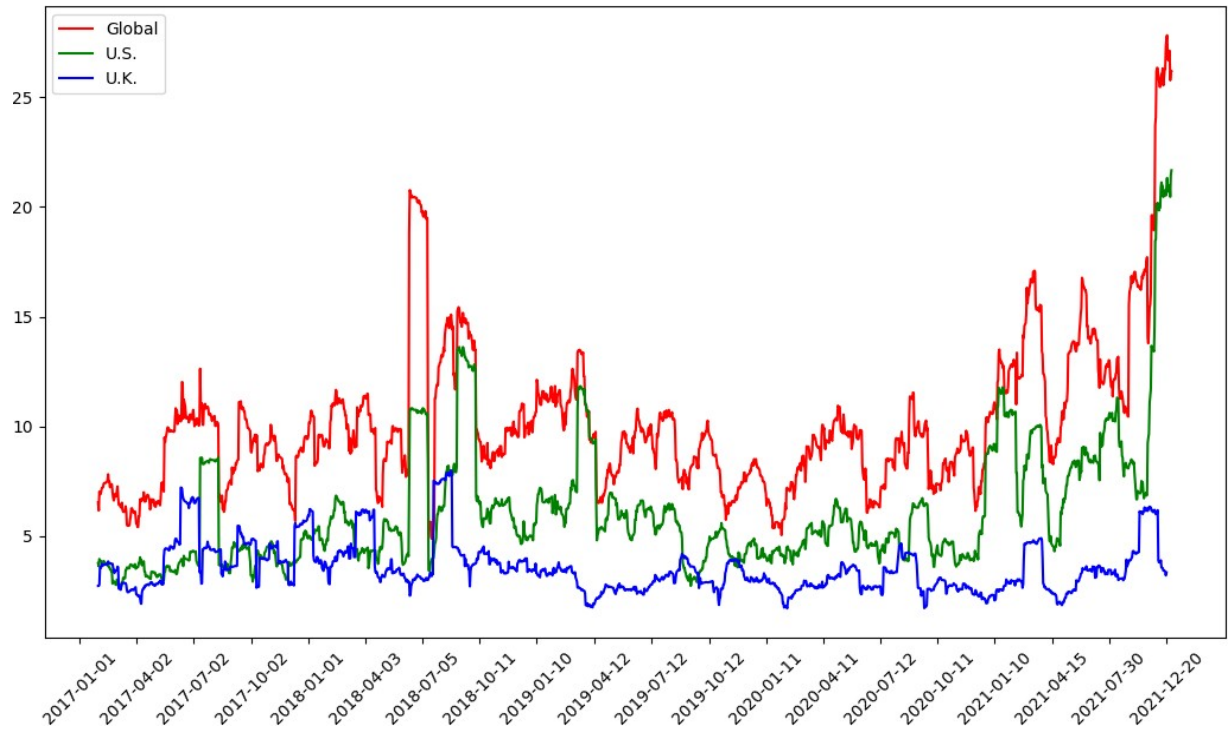

Figure 7. Global Sentiment Volatility

Table 6. Regression Result of Time-Series Models

|  |  | Before |  | During |  | After |  |
|---|---|---|---|---|---|---|---|
| Global | L(1) | 0.078*** | (0.01) | 0.36 | (0.49) | 0.157*** | (0.01) |
|  | L(2) | 0.064** | (0.04) |  |  |  |  |
|  | t | 8.20e-05*** | (0.00) | -1.58e-05 | (0.79) | -0.0001* | (0.08) |
|  | const | 0.129*** | (0.00) | 0.227*** | (0.00) | 0.190*** | (0.00) |
| U.S. | L(1) | 0.051* | (0.09) | 0.046 | (0.38) | 0.121** | (0.04) |
|  | t | 0.0001*** | (0.00) | 6.71e-07 | (0.99) | -0.0003** | (0.00) |
|  | const | 0.150*** | (0.00) | 0.245 | (0.00) | 0.226*** | (0.00) |
| U.K. | L(1) | 0.061** | (0.04) | 0.013 | (0.80) | -0.091 | (0.11) |
|  | t | 0.0001*** | (0.00) | -6.990e-05 | (0.69) | -9.48e-05 | (0.68) |
|  | const | 0.142*** | (0.00) | 0.255*** | (0.00) | 0.314 | (0.00) |

From Table 6, we observe that for global, U.S. and U.K., before the COVID-19 the coefficients of time trend are all significantly positive, indicating a steady positive attitude trend towards the inflation. During the COVID-19, the coefficients become insignificant, meaning that people are uncertain so that there is no clear trend. However, after COVID-19, the corresponding coefficients of global and U.S. scores are significantly negative, and the U.K. sentiment scores have negative coefficient, though not significant. Therefore, we can see people have a negative attitude trend after COVID-19. The visualized results are shown in Figure 8, Figure 9 and Figure 10.

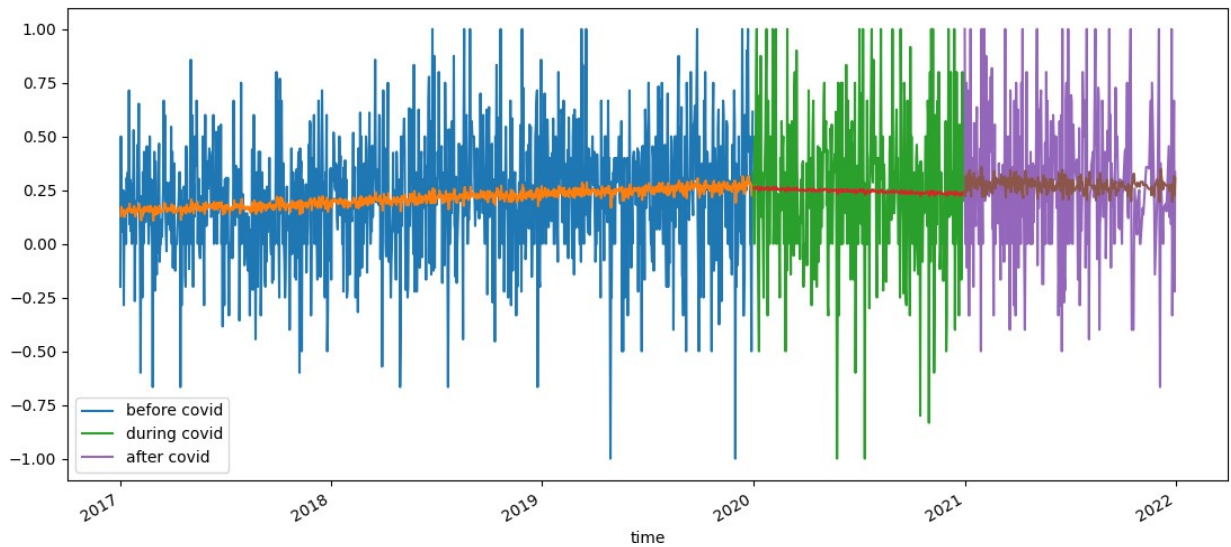

Figure 8. Trend of sentiment score of COVID-19 (Global)

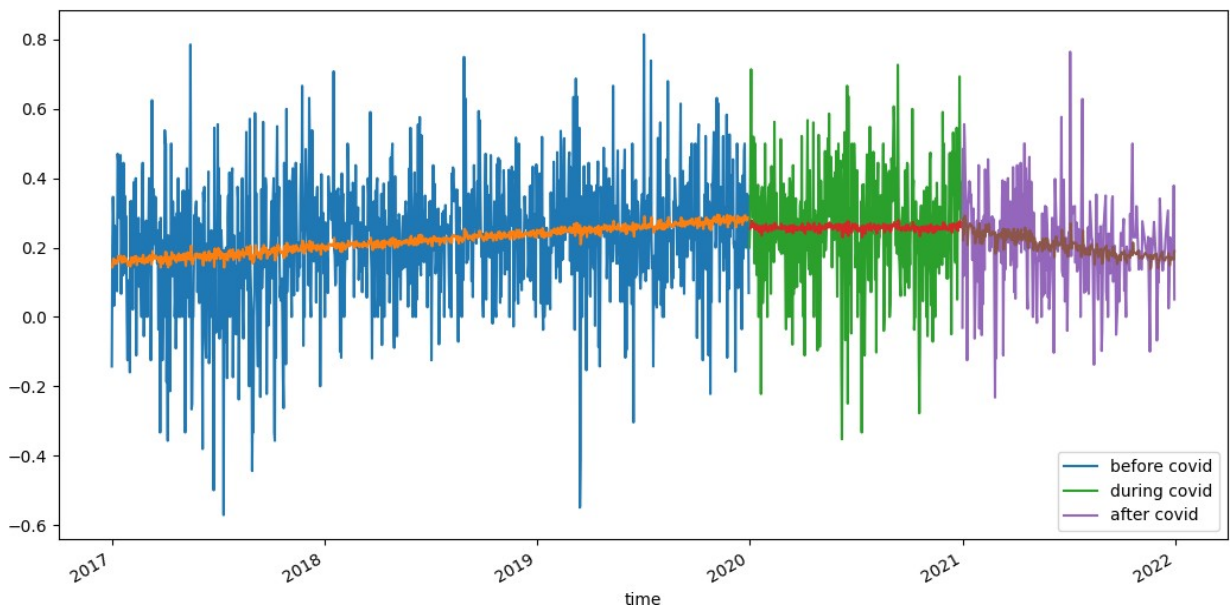

Figure 9. Trend of sentiment score of COVID-19 (U.S.)

Besides the expectation, we also measure the residuals, which is the signal of the level that people are panic about the inflation. The larger the mean square error is, the more panic people are because the more uncertainty of people's attitude there will be. We notice that the deviation of people's sentiment towards inflation changes in different episodes, which means that people are in more uncertain circumstances during or after COVID-19 than before COVID-19. We use the MSE (mean square error) of the regressions before to denote the level of people's uncertainty.

From Table 7 we observe that for U.S. and global sentiment, after COVID-19, the MSE becomes lower, indicating a universal negative sentiment towards the inflation. However, we observe an opposite trend in U.K., which means after COVID-19, British people hold more diversified opinion about the inflation.

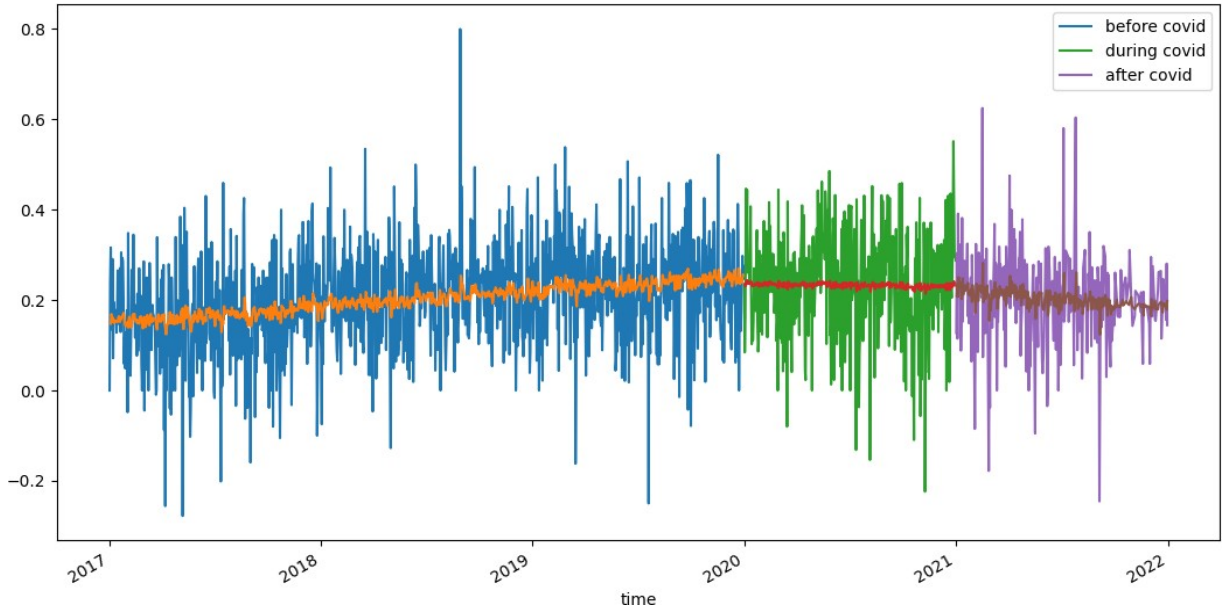

Figure 10. Trend of sentiment score of COVID-19 (U.K.)

Table 7. MSE of Regressions in Different Periods

|  | Before | During | After |
| --- | --- | --- | --- |
| Global | 0.012 | 0.015 | 0.009 |
| U.S. | 0.031 | 0.030 | 0.017 |
| U.K. | 0.080 | 0.121 | 0.111 |

## 5. Sentiment and Inflation

To analyze the relationship between people's sentiment sand inflation, we applied the data of Google trend, sentiment score of twitter attained above and the inflation data. Because the Google trend data is in weekly frequency and integer values, we took log for Google trend date and for sentiment data and inflation data, we calculated the weekly average. We used AR(2) model for the inflation rate and the regressions are as the following:

$$y_t = c_t + \varphi_1 y_{t-1} + \varphi_2 y_{t-2} + \varepsilon_t$$
$$y_t = c_t + \varphi_1 y_{t-1} + \varphi_2 y_{t-2} + \beta_1 Defl_{t-1} + \beta_2 Neu_{t-1} + \beta_3 Infl_{t-1} + \gamma_1 Senti_{t-1} + \varepsilon_t \quad (1)$$

The Defl, Infl, Neu represents log of lag-one sentiment data (Defl: deflation words, Infl: inflation words, Neu: neutral words), Senti represents the lagged sentiment scores, and y indicates the inflation rate in each country. We also used AR (2) model for inflation (L(1) and L(2)). Results of U.S. have been shown in Table 8 and results of U.K. have been shown in Table 9

For U.S., we observe that after taking the sentiment data into consideration, the coefficients of lagged inflation change from statistically significant to insignificant, meaning that sentiment data can help predict the future inflation, and the sentiment data contains relative information for predicting the future inflation, which corresponds to the empirical evidence of Angelico et al. (2022). What's more, we find that after COVID-19, the coefficient of inflation becomes significant, indicating that after COVID-19, people care more about the inflation and their sentiment towards the inflation makes influences on the market.

Table 8. U.S. Sentiment and Inflation

| US Infl | Total | | Before | | During | | After | |
|---|---|---|---|---|---|---|---|---|
| const | 0.024 | 4.763** | 0.057* | -2.251* | 0.098 | 14.193*** | 0.298** | 6.182*** |
|  | (0.215) | (0.022) | (0.074) | (0.070) | (0.124) | (0.000) | (0.020) | (0.004) |
| Defl |  | -0.291** |  | 0.143** |  | -0.342 |  | 0.058 |
|  |  | (0.010) |  | (0.024) |  | (0.159) |  | (0.577) |
| Infl |  | 0.383*** |  | 0.062 |  | -0.104 |  | 0.376*** |
|  |  | (0.000) |  | (0.195) |  | (0.332) |  | (0.000) |
| Neu |  | -0.564 |  | 0.748*** |  | -2.614*** |  | -1.108** |
|  |  | (0.209) |  | (0.005) |  | (0.000) |  | (0.022) |
| Senti |  | -0.665* |  | 0.142 |  | 0.012 |  | -0.447 |
|  |  | (0.080) |  | (0.491) |  | (0.987) |  | (0.195) |
| L(1) | 1.250*** | 0.095 | 1.201*** | -0.020 | 1.284*** | 0.104 | 0.977*** | 0.206* |
|  | (0.000) | (0.116) | (0.000) | (0.798) | (0.000) | (0.355) | (0.000) | (0.056) |
| L(2) | -0.262*** | 0.062 | -0.234*** | 0.010 | -0.356*** | 0.072 | -0.088 | -0.041 |
|  | (0.000) | (0.301) | (0.003) | (0.901) | (0.008) | (0.516) | (0.520) | (0.711) |

Table 9. U.K. Sentiment and Inflation

| UK Infl | Total | | Before | | During | | After | |
|---|---|---|---|---|---|---|---|---|
| Const | 0.039 | 2.276** | 0.195*** | 3.452*** | 0.335** | 2.807* | 0.129 | 4.667 |
|  | (0.251) | (0.023) | (0.005) | (0.000) | (0.038) | (0.068) | (0.153) | (0.237) |
| Defl |  | -0.653*** |  | 0.051 |  | 0.262 |  | -0.489* |
|  |  | (0.000) |  | (0.545) |  | (0.225) |  | (0.055) |
| Infl |  | 1.458*** |  | -0.270** |  | -0.222 |  | 1.637*** |
|  |  | (0.000) |  | (0.040) |  | (0.537) |  | (0.000) |
| Neu |  | -0.525*** |  | 0.174 |  | -0.082 |  | -1.233 |
|  |  | (0.008) |  | (0.269) |  | (0.779) |  | (0.152) |
| Senti |  | 0.075 |  | 0.335*** |  | 0.010 |  | -0.446 |
|  |  | (0.621) |  | (0.005) |  | (0.965) |  | (0.185) |
| L(1) | 1.304*** | -0.012 | 1.252*** | -0.097 | 1.200*** | 0.092 | 1.308*** | -0.056 |
|  | (0.000) | (0.808) | (0.000) | (0.212) | (0.000) | (0.514) | (0.000) | (0.589) |
| L(2) | 1.304*** | 0.038 | -0.316*** | -0.022 | -0.314** | 0.003 | -0.340*** | -0.029 |
|  | (0.000) | (0.426) | (0.000) | (0.783) | (0.020) | (0.986) | (0.016) | (0.796) |

## 6. Conclusion

In this paper, we conduct thorough investigation into the methodology of sentiment analysis (SA) and systematically categorize the approaches applied in the sentiment analysis domain. Based on the literature review which contains all sorts of practice in the SA field, we develop our own methodology to tackle our mission of SA on inflation afterCovid-19. We explore the topic via various media including Twitter comments, google trend and market-oriented inflation data with the aim to implement a comprehensive research of inflation sentiment. During the preliminary analysis of the data gathered, we withdraw fundamental features from the twitter comment such as geographical distinction and highly correlated keywords. From the google trend index data, we observe evident annually cyclical phenomenon and inter-country distinction in terms of index magnitude. Based on the market-oriented treasury inflation curve, we uncover similar trends and inter-country variation.

For further investigation, we test several models and algorithms including both traditional machine learning (NB, CNB, RF, SVM) and deep learning (BERT) on the tweet data. Combined with the test results, we conduct both theoretical and practical evaluation of the pros and cons of the methods tested and select BERT to extract features from the tweets. We calculate average daily sentiment score to construct sentiment index. Based on the sentiment index, we further apply auto regression (AR), auto correlation function (ACF) and partial autocorrelation function (PACF) to further analyze the public's sentiment trend in pre/during/post Covid periods. The regression results display significantly positive correlation of time trend before Covid, insignificant correlation during Covid, and negative correlation after Covid in terms of US, UK and globe. Meanwhile, the calculations of the deviation of people's sentiment towards inflation suggest inter-country and time-varying differences.

We take a step forward to pin down the relationship between public's sentiment and inflation. By integrating sentiment score of twitter and google trend into the AR (2) model, we gain insights that the addition of sentiment score render slagged inflation change from statistically significant to in-significant which suggests the inflation-predicting power within twitter sentiment score. Meanwhile, we observe inter-country distinctions and provide latent explanations behind them.

In conclusion, assessing inflation expectation in the social media perspective provides a novel way to elicit inflation expectation. Furthermore, the twitter-based sentiment index conveys valuable information whose inflation-predicting power outperforms traditional inflation prediction model. Also, compared with traditional macroeconomic model which contains country-specific factors and has long adjustment cycle and time-lagging features, twitter-based sentiment approach offers a country-universal and high-frequency methodology to tackle inflation-related issues.

**Notes**

Notes 1. https://www.websiterating.com/zh-CN/research/twitter-statistics/#chapter-1

Notes 2. https://www.ecb.europa.eu/press/key/date/2019/html/ecb.sp190711~6dcaf97c01.en.html

Notes 3. Obtain the data: https://data.mendeley.com/datasets/z9zw7nt5h2/1

Notes 4. Snowball algorithm: https://snowballstem.org/